\documentclass[twocolumn,nofootinbib,prd,floatfix,preprintnumbers,amsmath,amssymb,groupedaddress,superscriptaddress]{revtex4}
\usepackage{dsfont}
\usepackage{epsfig}
\usepackage{slashed}
\usepackage{bbold}

\begin{document}
\title{Composite Leptoquarks at the LHC}
\author{Ben Gripaios}
\email{gripaios@cern.ch}
\affiliation{CERN PH-TH, Geneva 23, 1211 Switzerland}
\preprint{CERN-PH-TH/2009-190}
\begin{abstract}
If electroweak symmetry breaking arises via strongly-coupled physics, the observed suppression of flavour-changing processes suggests that fermion masses should arise via mixing of elementary fermions with composite fermions of the strong sector. The strong sector then carries colour charge, and may contain composite leptoquark states, arising either as TeV scale resonances, or even as light, pseudo-Nambu-Goldstone bosons. The latter, since they are coupled to colour, get a mass of the order of several hundred GeV, beyond the reach of current searches at the Tevatron.  
The same generic mechanism that suppresses flavour-changing processes suppresses leptoquark-mediated rare processes, making it conceivable that the many stringent constraints may be evaded.
The leptoquarks couple predominantly to third-generation quarks and leptons, and the prospects for discovery at LHC appear to be good. As an illustration, a model based on the Pati-Salam symmetry is described, and its embedding in models with a larger symmetry incorporating unification of gauge couplings, which provide additional motivation for leptoquark states at or below the TeV scale, is discussed.
\end{abstract}
\maketitle
\section{Introduction}
Strong coupling provides a solution of the hierarchy problem of electroweak symmetry breaking (EWSB) that is natural in the literal sense. That is to say, we have already in Nature an example where a large hierarchy, to wit, that between the Planck ($10^{19}$ GeV) and proton (GeV) mass scales, arises as a result of the logarithmic running of the QCD coupling constant and the onset of strong coupling in the infra-red. 

But if strongly-coupled physics is to explain EWSB, we must solve the flavour puzzle: How can we generate fermion masses at the TeV scale, whilst suppressing dangerous higher-dimensional operators that contribute to flavour-changing processes by a scale of $10^{3}$ TeV or higher? And can we simultaneously understand 
the pattern of fermion masses and mixings?

One way to address these issues is to posit that fermion masses arise via mixing of elementary fermions with composite, fermionic operators of the strong sector, which feel EWSB \cite{Kaplan:1991dc}. The observed fermions of the Standard Model (SM) then arise as the light mass eigenstates. The operators of the strong sector may have large anomalous dimensions, in 
which case the hierarchy between the electroweak and Planck scales may re-surface, via the running, in the Yukawa couplings of the low-energy effective theory, giving an plausible origin for the fermion hierarchies, if not an explanation of their detailed structure. What is more, there is a natural suppression of flavour-changing processes for the light generations, since the light fermions are those that are least mixed with the composite fermions of the strong sector. To evaluate whether the suppression suffices, we need a calculable model, of which there are few; nevertheless, the observation that the required suppression can very nearly be achieved in models based on the AdS/CFT correspondence \cite{Gherghetta:2000qt} 
is already encouraging. By comparison, in models where fermion masses arise via elementary fermions coupling bi-linearly to bosonic operators of the strong sector that carry the gauge quantum numbers of the SM Higgs boson,
general arguments seem to suggest that a theory with the required suppression cannot exist \cite{Luty:2004ye}.

Here, I wish to argue that the more favoured class of models, where fermion masses arise via mixing, may give rise to interesting physics at the LHC in the form of leptoquarks. Now, like the mythical phoenix, leptoquarks are a theoretical idea that have already burnt themselves many times in the fire of experimental data, and readers of a certain age will, no doubt, despair at yet another incarnation. Nevertheless, I shall argue that not only is the proposed incarnation theoretically plausible, but also that the myriad negative searches may not be enough to burn it just yet. Happily, the LHC should soon do so, provided that a suitable search strategy can be devised.

My starting point is the simple observation that, in models of this type, the strong sector necessarily carries colour, as well as electroweak, charges. 
This is in stark contrast to the less-favoured models, where the strong sector need only contain a colour-singlet operator playing the r\^{o}le of the Higgs. 
What is more, given that models in the favoured class contain {\em fermionic} resonances with both types of charge (colour and electroweak), in the form of quark composites, it does not require a great leap of faith to imagine that they may possess {\em bosonic} resonances, at the TeV scale, with both colour and electroweak charges, that can couple bi-linearly to a quark and a lepton. The latter are, in short, leptoquarks. In the original $SU(3)$ model of  \cite{Kaplan:1991dc}, for example, where the composite fermions arise as technibaryons, the leptoquarks arise as technimeson states. 

These plausibility arguments for leptoquarks can be reinforced by appealing to the evidence for unification of the SM gauge couplings. Such a unification can be achieved (with precision comparable to that of the MSSM prediction) in the context of strong EWSB under the assumptions that the Higgs boson and right-handed top quark are both composite, and that the strong sector has a global symmetry corresponding to some unification group \cite{Agashe:2005vg}. Now, if the strong sector has such a symmetry, then its resonances fill out multiplets of that symmetry and, if unification is achieved by embedding the colour along with one or both of the electroweak groups ($SU(2)_L$ and $U(1)_Y$) in a simple group, then leptoquark states become even more likely.

The observation that unification scenarios give rise to leptoquark states is, of course, not new \cite{Pati:1974yy}. The extra ingredients here are twofold. The first is that when unification arises in this way, the leptoquark states automatically arise near the scale of electroweak symmetry breaking, and, {\em ergo}, within the reach of the LHC. The second is that in the usual scenarios, leptoquarks are coupled to SM fermions with a strength comparable to a gauge coupling, such that TeV leptoquark states would be in gross contradiction with many limits on rare processes. Here, by contrast, leptoquark couplings, at least those to light fermions, are highly suppressed.

It turns out that the leptoquark states need not even be as heavy as a TeV, since they may also arise as pseudo-Nambu Goldstone bosons associated with the breaking of electroweak symmetry. If they do so, then they should certainly be within reach of the LHC, whatever energy it finally reaches. 

So far, I have paid scant regard to the ills of leptoquarks, of which several are known. Most seriously, leptoquarks can mediate decay of the proton. 
One way to overcome this is to promote an accidental symmetry that prevents proton decay in the SM to a global symmetry of the strongly-interacting sector. Since the proton is the lightest observed baryon, baryon number symmetry is the obvious choice \cite{Agashe:2004ci}, but lepton number symmetry, or some combination of the two, may also suffice. In the example of \S \ref{ps}, I choose $3B+L$ (or equivalently, fermion number), which is enough to suppress the most dangerous processes, like $p\rightarrow e^+\pi^0$. Even then, we need to make sure that the couplings to the elementary sector (and within the elementary sector itself), which need not respect the global symmetry, do not cause further problems.

More difficult to overcome in the usual scenarios are the myriad constraints coming from leptoquark contributions to rare processes, such as those violating lepton family number.
But in models of the type discussed here, where fermion masses arise via (\ref{mix}), leptoquark-mediated processes exhibit a natural suppression via the very same mechanism that suppresses other flavour-changing processes. Namely, the light fermions (for which the constraints are strongest) are those least mixed with the strong sector.
In the absence of a calculable model, I will simply give a rough argument that this mechanism conceivably allows the existing constraints to be overcome, with two exceptions. The latter are the decays $\mu \rightarrow e \gamma$ and $\tau \rightarrow \mu \gamma$, mediated by a loop containing a leptoquark and a top quark. Consistency with these requires either that leptoquarks couple exclusively to quarks of one chirality, or that they do not couple to the top quark at all. 

Lastly, searches at the Tevatron have already ruled out leptoquarks with masses of up to two hundred GeV or so (the precise limits depending on whether leptoquarks are vectors or scalars, and their branching ratios for decay to different SM fermions). For generic resonances at the TeV scale, we need not worry, but if leptoquarks arise as light PNGBs, we should ask why they have not yet been observed. Even more worrisome is the concern that PNGB leptoquarks could end up with negative mass-squareds  (like a Higgs boson) and condense {\em in vacuo}, breaking colour. Happily, we shall see below that PNGB leptoquark mass squareds are positive-definite and that the masses are of the order of several hundreds of GeV. In a nutshell, the reason for the difference with the usual story where the Higgs arises as a PNGB \cite{Kaplan:1983fs} \cite{Agashe:2004rs,Contino:2006qr,Barbieri:2007bh} (and gets a {\em vev}) is that the dominant contributions to the PNGB leptoquark effective potential (and hence its mass) come from the QCD gauge coupling, rather than from couplings to the top quark, which dominate in the PNGB Higgs case.

The phenomenology is dominated by processes involving third generation quarks and leptons, to which these leptoquarks couple most strongly. Depending on their mass and electric charge, the leptoquarks will decay to either a top or a bottom quark, together with either a tau or a tau neutrino. There are also interesting possibilities for the observation of leptoquark-mediated rare processes, including $B\rightarrow K\mu \overline{\mu}$ , $\mu \rightarrow e \gamma$, $\tau \rightarrow \mu \gamma$, and $\mu-e$ conversion in nuclei, where my estimates for the leptoquark couplings, which may be considered as rough theoretical lower bounds, lie close to experimental upper bounds, either actual or envisaged.

The outline is as follows. In the next Section, I estimate the leptoquark contribution to various rare processes in this scenario and argue that the constraints can be evaded in principle. In \S \ref{light} I discuss the motivation for, and constraints from direct searches on, light PNGB leptoquarks. In \S \ref{ps}, I present an explicit model of PNGB leptoquarks, based on the Pati-Salam $SU(4)$ symmetry, which provides a concrete illustration of the general arguments. Phenomenological opportunities are discussed in \S \ref{pheno}.
\section{Rare Processes} \label{decay}
To estimate the leptoquark contribution to rare processes, one first needs an estimate of the various leptoquark couplings. In the scenario discussed here, these arise in much the same way as the usual SM Yukawa couplings of quarks and leptons to the composite Higgs. Namely, both types of coupling arise via mixing of elementary fermions, $q,u,d,l,e$ and perhaps $\nu$, with composite, Dirac-fermion operators of the strong sector $\mathcal{O}_{q,u,d,l,e,\nu}$, which in turn couple to composite Higgs and leptoquark states.

Let us follow \cite{Giudice:2007fh}, assuming that the strong sector is characterized by some resonance mass scale, $m_\rho$, and a coupling between the resonances, $g_\rho$, where we assume that $g_\rho$ is intermediate in strength between the SM couplings and the limit of $4\pi$ suggested by na\"{\i}ve dimensional analysis \cite{Manohar:1983md}. Then the up-type Yukawa coupling of the SM, for example, arises from the schematic Lagrangian \cite{Contino:2006nn,Giudice:2007fh}
\begin{multline} \label{mix}
\mathcal{L} \sim m_\rho \Big( \frac{y^q}{g_\rho} \overline{q} \mathcal{O}_{q} +   \frac{y^u}{g_\rho} \overline{u} \mathcal{O}_{u} +  \\ \overline{\mathcal{O}}_{q} \mathcal{O}_{q} + \overline{\mathcal{O}}_{u} \mathcal{O}_{u}\Big)  + g_\rho \overline{\mathcal{O}}_{q} H \mathcal{O}_{u}  +  h.c.,
\end{multline}
where $H$ represents the composite Higgs and I have written the couplings according to the rubric laid out in \cite{Giudice:2007fh}. 
The usual up-type SM Yukawa coupling is obtained in the low-energy effective theory by integrating out the resonances and is given in terms of the mixing parameters $y^{q,u}$ by
\begin{gather} \label{yuk}
\lambda^{u} \sim \frac{y^q  y^u}{g_\rho}.
\end{gather}
The leptoquark couplings arise in much the same way, with the coupling of a leptoquark to a quark, $Q \in \{ q,u,d \} $, and a lepton, $L  \in \{ l,\nu,e \}$, given by
\begin{gather} \label{lq}
\lambda^{LQ} \sim \frac{ y^L y^Q }{g_\rho}.
\end{gather}
 
Since there are more mixing parameters than measured SM Yukawa couplings, and since these relations are only approximate, it is clear that there is some room to man{\oe}uvre in trying to satisfy the constraints.
To argue that it can be achieved in principle, let me simply exhibit a set of mixing parameters that can reproduce the SM Yukawa couplings with $g_\rho \lesssim 4\pi$
whilst adequately suppressing leptoquark-mediated processes.\footnote{In a realistic model, one would also have to ensure that the whole gamut of flavour and electroweak precision constraints may be satisfied simultaneously, including contributions from non-leptoquark-mediated processes. This results in some tension even in the absence of leptoquarks. The $S$-parameter is an obvious example, but more pertinent for us are the $T$-parameter and the $Zb\overline{b}$ coupling. These may receive sizable contributions from the couplings $y^Q$, which are, therefore, subject to additional constraints. These latter constraints can, to some extent, be alleviated by judicious implementation of custodial symmetries, together with the introduction of light, fermionic, `custodian' resonances. See \cite{Agashe:2006at,Contino:2006qr,Contino:2006nn,Giudice:2007fh,Davidson:2007si} for more details.} To do so, I restrict to the simple case in which there just one mixing parameter for the quarks and one for the leptons of each generation, such that
\begin{gather} \label{favour1}
y^{q^i} \sim y^{u^i} \sim y^{d^i} , \; \; \forall \; i \in \{1,2,3 \},
\end{gather}
in the quark sector and
\begin{gather} \label{favour2}
y^{l^i} \sim y^{\nu^i} \sim y^{e^i}  , \; \; \forall \; i \in \{1,2,3\},
\end{gather}
in the leptonic sector, where the index $i$ labels the three generations. I then choose the smallest values of the mixing parameters that can reproduce the mass of the heaviest quark or lepton in each generation, ensuring that all the masses can be generated without excessive couplings in the strong sector. Thus one has 
\begin{gather}
y^{q^1} \sim  \sqrt{\lambda^d g_\rho},  \;  \; y^{q^2} \sim \sqrt{\lambda^c  g_\rho},  \;  \; y^{q^3} \sim \sqrt{\lambda^t  g_\rho},
\end{gather}
in the quark sector and
\begin{gather}
y^{l^1} \sim  \sqrt{\lambda^e  g_\rho}, \;  \;
y^{l^2} \sim \sqrt{\lambda^\mu  g_\rho}, \;  \;
y^{l^3}  \sim \sqrt{\lambda^\tau  g_\rho},
\end{gather}
in the lepton sector.

The leptoquark couplings in this scenario may then be estimated using (\ref{lq}). Using the upper endpoints of the fermion masses in Table \ref{SMmass} (which are given at a scale of 3 TeV in the $\overline{MS}$ scheme, with the one-loop running of quark masses due to QCD taken into account),
the resulting leptoquark couplings are as given in Table \ref{LQ}. Since these are, in some sense, the smallest possible leptoquark couplings consistent with the SM Yukawa couplings, they should be considered as lower bounds on the couplings in this scenario. Note that, following the assumptions, there is no distinction between the leptoquark couplings for the different quarks (or leptons) within a generation. 
\begin{table}
\begin{tabular}{| c | c | c |}
\hline
Fermion, $ f$ & Mass, $m_f$ & Yukawa, $\lambda^f$ \\
\hline
$e$ & $0.511$ MeV &$2.87 \times 10^{-6}$\\
$\mu$ & $106$ MeV &$6.09 \times 10^{-4}$\\
$\tau$ & $1.78$ GeV &$1.02 \times 10^{-2}$\\
\hline
$d$ & $2-4$ MeV &$2.30 \times 10^{-5}$\\
$s$ & $47 \pm 12$ MeV &$3.39 \times 10^{-4}$\\
$b$ & $2.40 \pm 0.04$ GeV &$1.40 \times 10^{-2}$\\
\hline
$u$ &$0.75-1.5$ MeV &$8.62 \times 10^{-6}$ \\
$c$ & $560\pm 40$ MeV &$3.43 \times 10^{-3}$\\
$t$ &$136 \pm 3$ GeV &$8.00 \times 10^{-1}$\\
\hline
\end{tabular}
\caption{SM fermion masses and Yukawa couplings corresponding to the upper limits. The masses are given at 3 TeV in the $\overline{MS}$ scheme, including one-loop QCD running for the quarks. Taken from \cite{Csaki:2008zd}. \label{SMmass}}
\end{table}
\begin{table}
\begin{tabular}{| c | c | c | c |}
\hline
Lepton $\backslash$ Quark  & 1 &2& 3 \\
\hline
1  & $ 8.2\times 10^{-6}$ &$1.0\times 10^{-4}$& $1.5\times 10^{-3}$ \\
2  & $1.2\times 10^{-4}$ &$1.5\times 10^{-3}$& $2.2 \times 10^{-2}$ \\
3  & $4.9\times 10^{-4}$ &$6.0 \times 10^{-3}$& $9.0\times 10^{-2}$ \\
\hline
\end{tabular}
\caption{Estimates of leptoquark couplings, using fermion masses in Table \ref{SMmass} and assumptions described in the text. \label{LQ}}
\end{table}

With estimates for the leptoquark couplings in hand, one can compare with the existing experimental constraints. Since the estimates of the leptoquark couplings are so crude, a rough comparison with the data will suffice; one could of course attempt a more rigorous comparison given a specific (and calculable) holographic model, for example. 

To compare, I use a leptoquark mass of 225 GeV, near the Tevatron bound, though it should be borne in mind that the leptoquarks are expected to be somewhat heavier in this scenario, even if they arise as PNGBs; in any case, the bounds scale inversely with the square of the leptoquark mass.

For processes not involving the third generation, all constraints are avoided by several orders of magnitude. For example, the ratio $\frac{\Gamma (\pi \rightarrow e\nu) }{\Gamma (\pi \rightarrow \mu \nu)}$, which exhibits a chiral suppression in the SM, places strong constraints on non-chirally coupled leptoquarks.  A recent analysis \cite{Campbell:2008um} gives the following constraints on couplings of vector\footnote{The bounds for scalar leptoquarks are weaker by a factor of two typically.} leptoquarks of mass 225 GeV:
\begin{align}
\lambda^{11}_L \lambda^{11}_R &< 1.2 \times 10^{-6}, \nonumber \\
\lambda^{11}_L \lambda^{21}_R &< 6.6 \times 10^{-3}, \nonumber \\
\lambda^{21}_L \lambda^{11}_R &< 1.5 \times 10^{-5}.
\end{align}
In comparison, I estimate the largest contribution to $\lambda^{11} \lambda^{11}$ to be below $10^{-10}$, while the estimate for $\lambda^{21} \lambda^{11}$ is of order $10^{-9}$, both four orders of magnitude smaller.
Similarly, one can consider the conversion of muons to electrons in scattering off nuclei \cite{Kuno:1999jp}. Applying the analysis of \cite{Davidson:1993qk} to the best limit (which comes from a conversion rate of less than $7\times 10^{-13}$ in Gold \cite{Bertl:2006up}), we find a bound
$\lambda^{11} \lambda^{21} < 2 \times 10^{-6}$;
the estimate is over three orders of magnitude lower. 

The constraints tighten if one considers processes involving both quarks and leptons of the second generation. For $K\rightarrow \overline{\mu}e$,
a bound of $\lambda^{11}\lambda^{22} < 1.5 \times 10^{-6}$ is given in \cite{Dreiner:2001kc}, to be compared with my estimate of $10^{-8}$. $K\rightarrow \overline{\mu} \mu$ should give a stronger constraint, since an electron from the first generation is replaced by a muon from the second generation, but is a little harder to treat, since it receives a non-zero contribution from the SM and indeed is observed with branching ratio $6.84 \pm 0.11 \times 10^{-9}$ \cite{Amsler:2008zzb}. To get a rough bound, I follow \cite{FileviezPerez:2008dw}, requiring that the leptoquark-mediated contribution
be at most comparable to the theory error in the SM contribution, $\Delta B < 0.6 \times 10^{-9}$. By na\"{\i}vely scaling the above bound from $K\rightarrow \overline{\mu}e$, I get the bound $\lambda^{21}\lambda^{22} < 1.5 \times 10^{-5}$, which is a hundred times my estimate.

The most interesting processes involve one or more third-generation quarks or leptons. A bound on $\lambda^{22}\lambda^{32} < 4.7 \times 10^{-3}$, from $\tau \rightarrow \eta \mu$ data collected up to 2006 \cite{Dreiner:2006gu}, is evaded by a few hundred and a bound on $\lambda^{22}\lambda^{23} < 5.0 \times 10^{-3}$,
from $B_s \rightarrow \overline{\mu} \mu$  \cite{Dreiner:2006gu}, by a hundred or so. Similarly, safety factors of at least fifty are obtained if one considers the decays $K\rightarrow \pi \overline{\nu} \nu$, $B \rightarrow \tau \overline{\mu} X$, or $B \rightarrow \tau \overline{\tau} X$ \cite{Davidson:1993qk}. 

However, there are three processes for which the estimates lie tantalizingly close to, or even exceed, the bounds. The first of these is $B_d \rightarrow K \mu \overline{\mu}$, for which the bound \cite{Xu:2006vk} is $\lambda^{22} \lambda^{23} < 2.3 \times 10^{-4}$,
compared to my estimate of $3 \times 10^{-5}$. The other two processes are the radiative lepton decays $\mu \rightarrow e \gamma$ and $\tau \rightarrow \mu \gamma$, which can be mediated by a loop containing a leptoquark and a quark of the third generation. The bounds here depend on whether or not the leptoquarks are chiral, meaning that they couple to quarks of just one chirality or not. For chiral leptoquarks, on the one hand, the helicity flip necessary for $l\rightarrow l^\prime \gamma$ must be on an external fermion. To make a rough estimate, I use the formula \cite{Davidson:1993qk}
\begin{gather}
\Gamma_{l\rightarrow l^\prime \gamma} \sim 2e^2 \left( \frac{m_l}{8\pi}\right)^5 \left(\frac{\lambda^{l3} \lambda^{l^\prime 3}}{m_{lq}^2} \right)^2,
\end{gather}
valid for $m_l \gg m_{l^\prime}$ and $m_{lq} \gg m_{b,t}$, up to factors of order unity coming from the electric charges of the quarks and leptoquarks. Using the MEGA bound of $BR(\mu  \rightarrow e \gamma) < 1.2 \times 10^{-11}$ \cite{Ahmed:2001eh} and the Belle/BABAR bounds of $BR(\tau  \rightarrow \mu \gamma) < 4.5 \times 10^{-8}$ \cite{Hayasaka:2007vc,Aubert:2005ye}, gives bounds of $\lambda^{13} \lambda^{23} < 2.0 \times 10^{-4}$ and $\lambda^{23} \lambda^{33} < 2.9 \times 10^{-2}$, respectively.
These are just a factor of ten or so above my estimates of $3 \times 10^{-5}$ and $2 \times 10^{-3}$, respectively. 

On the other hand, for non-chirally coupled leptoquarks, the helicity flip can be moved to the internal quark, with a resulting enhancement of $\frac{m_{t,b}}{m_l}$ of the above bounds. For processes mediated by the top, this certainly violates the bound, and so it must be the case that the leptoquarks are chiral, or that they do not couple to the top. A consultation of the possible leptoquark states and their possible couplings, given in \cite{Davidson:1993qk}, shows that this restriction is not so severe. 

Even though my lower-bound estimates are so close to the experimental upper-bounds for these three processes, even for chiral leptoquarks, there are several reasons why one should not hope for either an imminent discovery or an exclusion of this scenario via indirect searches. Firstly, as we shall see in \S \ref{pheno}, the prospects for significant improvements in the experimental bounds are not great. Secondly, my estimates are coarse and should be treated with circumspection, if not outright derision. Thirdly, the estimates have been derived for leptoquark masses at the Tevatron bound of 225 GeV, whereas the natural expectation for leptoquark masses in this scenario is closer to a TeV. In summary, we see that it is conceivable that the many constraints can be evaded in models of this type. 
\section{Light Leptoquarks} \label{light}
Thus far, I have only argued that the leptoquarks arise as resonances at roughly the TeV scale; in the absence of a concrete (and calculable) model, one cannot say for sure whether they will be in reach of the LHC. However, there is one scenario in which leptoquarks are naturally lighter, which occurs when they arise as pseudo-Nambu Goldstone bosons (PNGBs) of some spontaneously-broken, approximate symmetry. What is more, such PNGBs lie within the regime of validity of the sigma model describing the low-energy effective field theory, such that one can make quantitative statements about the dynamics.

That PNGBs of this type can arise should hardly come as a surprise. This is, after all, how the pions and kaons arise in our current understanding of QCD, and, moreover, we already know that a strong sector that is responsible for EWSB must have at least three NGBs, furnishing the longitudinal components of the $W^\pm$ and $Z$ gauge bosons. Whether or not further PNGBs arise depends, of course, on the pattern of symmetry breaking. As examples, the minimal composite Higgs model \cite{Agashe:2004rs,Contino:2006qr}, based on the coset $SO(5)/SO(4)$, features a fourth PNGB that plays the role of the Higgs, whereas a model based on $SO(6)/SO(5)$ \cite{Gripaios:2009pe} contains, in addition, an electroweak singlet $\eta$ that can be used to hide the Higgs from LEP via the dominant decay $ h \rightarrow \eta \eta$, as well as the possibility of new physics associated with anomalies \cite{Gripaios:2008ei}. In all these examples, one assumes that the $SU(3)$ colour symmetry factorizes, but it need not do so. If it does not, then light leptoquarks may arise as PNGBs.

If leptoquarks are light, there is an immediate worry that the masses of PNGB leptoquarks, which arise via gauge and fermion couplings that break the symmetry explicitly, may be too small, such that PNGB leptoquarks would have been detected at existing colliders such as the Tevatron. Even worse, there is the worry that the squared masses come out negative, like that of the Higgs, in which case colour would not be a symmetry of the vacuum. Happily, such worries are unfounded: Unlike the PNGB Higgs, for which the dominant contribution to the effective potential (and the mass) comes from the top quark Yukawa coupling and is negative, the dominant contribution to the squared mass of a PNGB leptoquark comes from the strong interaction, and is positive. Indeed, NDA suggests that $m_{LQ}^2 \sim g_\rho^2 f^2\frac{\alpha_s}{4\pi} $, where $f \gtrsim v = 246$ GeV is the analogue of the pion decay constant in the strong EWSB sector. So even for a model without a Higgs, for which $f=v$, we expect a PNGB leptoquark mass of order a few hundred GeV. For models with a PNGB Higgs and $v < f$, which make it easier to fit the electroweak precision data, we can expect a mass more like 700 GeV (for $f \sim 500 GeV$). 

Since the leptoquark couplings scale with the fermion masses, one expects decays to third generation decays (to top quarks in particular if allowed)  to dominate. 
By comparison, searches at D0 for third generation scalar leptoquarks decaying to $b\tau$ \cite{Abazov:2008jp} or $b\nu_\tau$ \cite{Abazov:2007bsa} yield bounds of $210$ and $229$ GeV, respectively. 

I should stress that the idea that light leptoquarks might arise as PNGBs is not new \cite{Schrempp:1984nj}. What is new is the natural suppression of rare decays mediated by leptoquarks when fermion masses arise via mixing as in (\ref{mix}).\footnote{In \cite{Schrempp:1984nj}, rare decays were suppressed because leptoquarks only coupled to fermions via dimension-five operators involving a derivative. Unlike the suppression mechanism discussed here, this is not generic.}

\section{The Pati-Salam Model} \label{ps}
In what follows, I would like to illustrate the foregoing general ideas by means of a simple example, with scalar leptoquarks arising as PNGBs. The possible choices of the symmetry breaking structure that gives rise to the leptoquark states are almost endless; for concreteness, I focus on the example described in \cite{Schrempp:1984nj}, namely where $SU(3)_c$ and the $B-L$ part of the $U(1)_Y$ are embedded in the $SU(4)_{PS}$ of Pati and Salam \cite{Pati:1974yy}. In order to prevent rapid decay of the proton, the strong sector is further endowed with an additional $U(1)_{3B+L}$ symmetry.\footnote{Even with this symmetry, nucleons may decay, {\em e.g.}\ $n \rightarrow 3\nu$, via operators of dimension nine or higher. Amusingly, the best bound on a $(3B+L)$-preserving decay comes from attributing the observed muon neutrino flux to decays of all the Earth's baryons, giving $\tau (n \rightarrow 3\nu_\mu) < 5 \times 10^{26}$ yr \cite{Learned:1979gp}, with an effective suppression scale of a few hundred TeV. Since this process involves six light fermions, for which the elementary/composite mixing can be small, the required suppression is achievable. I thank K. Agashe for bringing this issue to my attention.}
 
The rest of the SM gauge group goes into at least an $SU(2)_L \times SU(2)_R$ subgroup. Beyond that there are a number of possiblities. In order to get a Higgs boson as a PNGB, one might put the $SU(2)_L \times SU(2)_R$ into either an $SO(5)_{LR}$ \cite{Agashe:2004rs,Contino:2006qr} or an $SO(6)_{LR} \cong SU(4)_{LR}$ \cite{Gripaios:2009pe}. To achieve grand unification, one might then imagine either putting $SU(4)_{PS} \cong SO(6)_{PS}$ with $SO(5)_{LR}$ or $SO(6)_{LR}$ to make $SO(11)$ or $SO(12)$, or just adding a discrete parity interchanging $SU(4)_{PS} \leftrightarrow SU(4)_{LR}$. Given the rather large rank of these groups, one might fear a slew of PNGB leptoquarks, but this is not necessarily the case. A model based on $SO(11)/SO(10)$, for example, contains as PNGBs just a single SM Higgs boson plus a single, complex scalar leptoquark
with charges $(\mathbf{3},\mathbf{1},-\frac{1}{3})$ under $SU(3)_c \times SU(2)_L \times U(1)_Y$.

Let me focus on the Pati-Salam example, assuming that $SU(4)_{PS}$ is broken to $SU(3)_c \times U(1)_{B-L}$ by the same dynamics that drives EWSB. This will also allow us to compare and contrast with  \cite{Schrempp:1984nj}. The PNGB leptoquarks then correspond to generators in the coset $SU(4)_{PS}/SU(3)_c \times U(1)_{B-L}$. There are six degrees of freedom, corresponding to a complex, colour-triplet, scalar, with charges $(\mathbf{3},\mathbf{1},\frac{2}{3})$ under $SU(3)_c \times SU(2)_L \times U(1)_Y$, and uncharged under $U(1)_{3B+L}$. They couple to quarks and leptons (namely $\overline{q_L} l_L$, $\overline{u_R} \nu_R$ and  $\overline{d_R} e_R$) only in dimension five operators, containing a derivative.

It is also necessary to specify the composite fermions to which the elementary fermions couple. As in the usual set-up of Pati and Salam, the $\mathbf{4}$ contains states with  the required SM gauge charges. Explicitly, one can couple left-handed elementary fermions (which mix to give the SM fermions $q_L$ and $l_L$) to composite Dirace fermions in the $(\mathbf{4},\mathbf{2}, \mathbf{1}, 1)$ of $SU(4)_{PS} \times SU(2)_L \times SU(2)_R \times U(1)_{3B+L}$, and right-handed elementary  fermions (which mix to give $u_R, d_R, e_R$, and $\nu_R$ if desired) to $(\mathbf{4},\mathbf{1}, \mathbf{2},1)$s. In passing, it is worth remarking that the couplings between elementary and composite sectors do not respect the full global symmetry of the strong sector. In particular, they need not respect $U(1)_{3B+L}$, with potentially dangerous consequences for proton decay. But in the case at hand, the SM gauge symmetries (which certainly are respected by the elementary/composite couplings) guarantee that only left-handed elementary fermions can couple to composite $(\mathbf{4},\mathbf{2}, \mathbf{1}, 1)$s. Conversely, only right-handed elementary fermions couple to the $(\mathbf{4},\mathbf{1}, \mathbf{2}, 1)$s. Since $q_L$ and $l_L$ both have $3B+L = +1$, for example, one sees that $U(1)_{3B+L}$ is preserved as an accidental symmetry by these couplings.\footnote{I thank K. Agashe for discussion on this point.} 

To write down the sigma model for the PNGB leptoquarks, choose a basis of generators $T^{i} = \frac{\lambda^{i}}{\sqrt{2}}$ for the fundamental of $SU(4)$ as described in the Appendix. In that basis, $SU(4)$ is broken to $SU(3)_c\times U(1)_{B-L} $, generated by $\lambda^{1\dots 8}$ and $\lambda^{15}$, via a field $\Sigma$ transforming in the adjoint of $SU(4): \Sigma \rightarrow U \Sigma U^\dagger$, with {\em vev}
\begin{gather}
\Sigma_0 = \frac{1}{2\sqrt{2}} \begin{pmatrix} 1 & 0 \\ 0 & -3 \end{pmatrix},
\end{gather}
where the upper-left matrix entry denotes the $3\times 3$ unit matrix. Define the $\Sigma$-model field by
\begin{gather}
\Sigma = e^{i\frac{\Pi}{2f}} \Sigma_0 e^{-i\frac{\Pi}{2f}},
\end{gather}
where
\begin{gather}
\Pi = \begin{pmatrix}  0 & \chi \\ \chi^\dagger& 0\end{pmatrix},
\end{gather}
the upper-left entry denotes the $3\times 3$ zero matrix, and $\chi$ is a column vector with three entries. Then,
\begin{multline}
\Sigma = \Sigma_0 +\frac{i}{ \sqrt{2\chi^\dagger\chi}}\sin{\frac{\sqrt{\chi^\dagger\chi}}{f}}\begin{pmatrix} 0 & -\chi \\ \chi^\dagger& 0\end{pmatrix} 
\\ + \frac{1}{\sqrt{2}\chi^\dagger\chi}\Big(\cos{\frac{\sqrt{\chi^\dagger\chi}}{f}} -1\Big)
 \begin{pmatrix}  \chi \chi^\dagger& 0 \\ 0& \chi^\dagger\chi\end{pmatrix}.
\end{multline}
To clean things up slightly, one may re-define \cite{Coleman:1969sm}
\begin{gather}
\phi =  \chi \frac{ f}{\sqrt{\chi^\dagger\chi}}\sin{\frac{\sqrt{\chi^\dagger\chi}}{f}},
\end{gather}
in terms of which
\begin{gather}
\Sigma =\Sigma_0 + \frac{i}{ \sqrt{2}f}\begin{pmatrix} 0 & -\phi \\ \phi^{\dagger}& 0\end{pmatrix}  \\+\frac{1}{ \sqrt{2}}\Big(\sqrt{1-\frac{\phi^{\dagger}\phi}{f^2}}-1\Big)  \begin{pmatrix}  \frac{\phi \phi^{\dagger}}{\phi^{\dagger}\phi}& 0 \\ 0& 1\end{pmatrix}.
\end{gather}
\subsection{The leptoquark effective potential}
Now let me compute the one-loop, Coleman-Weinberg, effective potential \cite{Coleman:1973jx} for the leptoquarks. The largest contributions will come from the coupling to QCD and the couplings to third generation quarks.

Firstly, write the general low-energy effective Lagrangian at quadratic order in the fields, obtained by integrating out the heavy degrees of freedom of the strong sector, as
\begin{widetext}
\begin{multline} \label{lag}
\mathcal{L} = \frac{1}{2}P_{\mu \nu} \Big[\Pi_0^A (p^2) \mathrm{tr}  \; A^{\mu} A^{\nu} + \Pi_1^A (p^2)\mathrm{tr} \; | i[ A^\mu, \Sigma] |^2\Big] 
\\+ \Sigma_{i,j\in \{q_L,l_L \} } \overline{\Psi}_i \slashed{p}\Big[\Pi_0^{ij} (p^2) +  \Pi_1^{ij} (p^2) \Sigma\Big] \Psi_j + \Sigma_{k,l\in \{ u_R,d_R,e_R \} } \overline{\Psi}_k \slashed{p} \Big[\Pi_0^{kl} (p^2) +  \Pi_1^{kl} (p^2) \Sigma^\dagger \Big] \Psi_l,
\end{multline}
where $\Pi_{0,1}$ are unknown form factors, and I have used an $SU(4)$-covariant notation. The largest contributions come from QCD, for which the gauge field is 
$A^\mu =  \begin{pmatrix} a^\mu & 0 \\ 0 & 0\end{pmatrix}$
and from the couplings to $q_L$ and $t_R$. These are encompassed in the gauge terms
\begin{gather}
\mathcal{L} = \frac{1}{2}P_{\mu \nu} \Bigg[\Pi_0^A (p^2) \mathrm{tr} \; a^{\mu} a^{\nu} + \Pi_1^A (p^2)\Big(2(1-\sqrt{1-\frac{\phi^{\dagger}\phi}{f^2}})\frac{\phi^{\dagger}a^{\mu 2} \phi}{\phi^{\dagger} \phi}- \big[(1-\sqrt{1-\frac{\phi^{\dagger}\phi}{f^2}}) \frac{\phi^{\dagger}a^{\mu } \phi}{\phi^{\dagger} \phi}\big]^2\Big)\Bigg],
\end{gather}
and in the fermion terms
\begin{gather}
\mathcal{L} = \Sigma_{r=t_L,b_L,t_R} \overline{r} \Big[\Pi_0^{rr} (p^2) +  \Pi_1^{rr} (p^2)\frac{1}{\sqrt{2}} \big(1-\sqrt{1-\frac{\phi^{\dagger}\phi}{f^2}}\big)  \frac{\phi\phi^{\dagger}}{\phi^{\dagger} \phi}\Big] r.
\end{gather}

To compute the one-loop Coleman-Weinberg effective potential, use the $SU(3)_c$ symmetry to set $\phi_i = \epsilon f \delta^3_i$, with $0 \leq \epsilon \leq 1$, such that
\begin{gather}
\mathcal{L} = \frac{1}{2}P_{\mu \nu} \Big[\Pi_0^A (p^2) \mathrm{tr}  \; a^{\mu} a^{\nu} + \Pi_1^A (p^2)\big(1-\sqrt{1-\epsilon^2}\big)\Sigma_{i=4,5,6,7} (a^{\mu i })^2
+\frac{2}{3} \epsilon^2 (a^{\mu 8})^2\Big],
\end{gather}
whence
\begin{gather}
V_g = \frac{3}{2} \int \frac{d^4 p}{(2 \pi)^4} \Bigg[4\log{\Big(\Pi_0^A + \Pi_1^A\big(1-\sqrt{1-\epsilon^2}\big)\Big)} + \log{(\Pi_0^A + \frac{2}{3} \epsilon^2 \Pi_1^A)}\Bigg].
\end{gather}
Assuming that $\Pi_1 \ll \Pi_0$, this becomes
\begin{gather}
V_g \simeq \alpha\Big[ \epsilon^2 + 6 (1-\sqrt{1-\epsilon^2})\Big],
\end{gather}
where $\alpha = \int \frac{d^4 p}{(2 \pi)^4} \frac{\Pi_1^A}{\Pi_0^A}$; this contribution favours a minimum at the origin. 

Similarly, from the fermion couplings we get
\begin{gather}
V_f = - \frac{4}{2}\int \frac{d^4 p}{(2 \pi)^4}\Bigg[2 \log{p^2\Big(\Pi_0^q + \frac{1}{\sqrt{2}}\Pi_1^q \big(1-\sqrt{1-\epsilon^2}\big)\Big)} +  \log{p^2\Big(\Pi_0^t + \frac{1}{\sqrt{2}}\Pi_1^t (1-\sqrt{1-\epsilon^2})\Big)}\Bigg].
\end{gather}
\end{widetext}
Assuming that $\Pi_1 \ll \Pi_0$, this becomes
\begin{gather}
2\beta (1-\sqrt{1-\epsilon^2}),
\end{gather}
where $\beta = \int \frac{d^4 p}{(2 \pi)^4}  \Big[ 2 \frac{\Pi_1^q}{\Pi_0^q}+  \frac{\Pi_1^t}{\Pi_0^t}\Big]$. Again, a minimum at the origin is favoured.
 
There are a number of points in which our calculation differs from that of the familiar calculation of the effective potential of a composite Higgs model \cite{Agashe:2004rs}, which lead to significant difference between the PNGB Higgs and PNGB leptoquarks. Firstly, for the Higgs the dominant contribution comes from the top quark, which favours a minimum away from the origin. This is partly cancelled by the leading sub-dominant contribution from the $SU(2)_L$ gauge field, which prefers a minimum at the origin. The result is that the Higgs gets a vacuum expectation value, but the mass of the Higgs is rather light. For the leptoquarks, the dominant contribution comes from QCD, and favours a minimum at the origin, so there is no danger that colour will be broken in the vacuum. Moreover, the leading sub-dominant contribution comes from top quarks, but it too favours a minimum at the origin, at least in the Pati-Salam example.

These arguments are reinforced by $N_c$ effects. For the Higgs, the quark contributions are  enhanced by $N_c$, while the $SU(2)_L$ gauge contributions are unenhanced. For the leptoquark, the gauge contribution is enhanced by the quadratic Casimir ($\frac{N_c^2 -1}{N_c}$) while the quark contributions are not enhanced.

As a result, we need not fear that the vacuum will break colour in a model with PNGB leptoquarks of this type. What is more, the PNGB leptoquark mass is expected to be rather large. An NDA estimate (assuming $y_{L,R}^t \sim \sqrt{\lambda_t g_\rho}$ as above) yields $m_{lq}^2 \sim (\frac{N_c^2 -1}{2N_c} 4\pi  \alpha_s + \lambda_t g_\rho )\frac{g_\rho^2}{16\pi^2} f^2$, , so taking $f \sim 500$ GeV as suggested by EWPT gives a PNGB leptoquark mass as high as a TeV, for $g_\rho \sim 4 \pi$.
\section{Phenomenology} \label{pheno}
The couplings in (\ref{lag}) also give rise to the couplings of a leptoquark to a lepton and a fermion, of the form
\begin{gather}
\mathcal{L} = \frac{-i}{\sqrt{2}f} \Pi_1^{ql} \overline{q}_L \slashed{p} \phi l_L + \frac{-i}{\sqrt{2}f} \Pi_1^{de} \overline{d}_R  \slashed{p} \phi e_R.
\end{gather}
A leptoquark $\phi$, produced at the LHC, can thus decay either to an up-type quark together with a neutrino, or to a down-type quark together with a charged lepton.
Now, there are two effects that essentially guarantee that the leptoquarks will decay almost exclusively to third generation fermions. The first effect, which is generic to letpoquark scenarios of this type, is simply that the leptoquark couplings increase with the SM Yukawa couplings. For the models studied in \S \ref{decay} for example, which are most favourable from the point of view of evading constraints from rare decays, the leptoquark couplings scale as the product of the square roots of the relevant quark and lepton Yukawa couplings. The second effect is not generic, but arises only in models, such as the Pati-Salam model just discussed, where the leptoquark couples to quarks and leptons via a derivative. 

As a result of the derivative in the coupling, there is an additional suppression in the decay to light fermions. Indeed, for $m_1, m_2 \ll m_{lq}$, one finds
\begin{gather}
\Gamma \sim |\Pi|^2 \frac{m_{lq} m_{1,2}^2}{32 \pi f^2}, 
\end{gather}
where $m_{1,2}^2$ represents some linear combination of $m_{1}^2$ and $m_{2}^2$, depending on how the derivative appears in the coupling.

As a result of either (or both) of these effects, one expects that any leptoquark in this scenario will decay predominantly to the heaviest possible quark and lepton. Furthermore, since our NDA estimates (and Tevatron observations) put the mass of any leptoquark above that of the top quark, we can expect decays involving the top to dominate. The associated lepton will either be a tau, or a neutrino, depending on the electric charge of the leptoquark. 
It is also possible, of course, that a generic leptoquark does not couple to the top quark, in which case one may expect decays to a $b$ quark, together with either a tau or a neutrino.

Since these leptoquarks couple dominantly to third-generation quarks and leptons, pair production through colour gauge interactions will overwhelmingly dominate single production at the LHC; cross-sections are given in \cite{Kramer:2004df}. The channels of interest are, therefore, $2t2\tau$, $2t + \slashed{E}_T$, $2b 2\tau$, and $2b + \slashed{E}_T$. Of these, the latter two have already been the subject of searches at the Tevatron, and presumably these could be adapted easily enough for the LHC, although I am aware of only one existing study \cite{Mitsou:2004hm}, suggesting a reach of up to $1.5$ TeV in mass in the $2b + \slashed{E}_T$ channel. The use of novel kinematic variables such as $m_{T2}$ \cite{Lester:1999tx} in this missing energy channel may well improve the prospects for discovery \cite{Barr:2009wu} and mass measurement. 
The two channels involving the top will require more ingenuity, but merit investigation.

As regards indirect searches, we saw earlier that the most promising channels were in $B_d \rightarrow K \mu \overline{\mu}$,  $\mu \rightarrow e \gamma$ and $\tau \rightarrow \mu \gamma$. The first of these, as well as $B_s \rightarrow \mu \mu$, will be probed at LHCb. For $\mu \rightarrow e \gamma$, the MEG experiment \cite{Cei:2009zz}, currently in operation, hopes to lower the bound on the branching ratio to $10^{-13}$, which is at the level of my estimate in \S \ref{decay}.
For $\tau \rightarrow  \mu \gamma$, a current or future $B$-factory \cite{Browder:2008em} offers the best hope \cite{Akeroyd:2004mj}, with sensitivity around $10^{-9}$ though CMS and ATLAS at the LHC may also be sensitive \cite{Serin}.

Bounds on other processes may also become relevant in the light of future data. For example, an intense muon source may yield a significant lowering of the bound on $\mu-e$ conversion in nuclei. The best hope seems to be the PRISM facility, which may reach a sensitivity of $10^{-19}$ \cite{Roberts:2005ux}. Alternatively, the mu2e proposal \cite{Carey:2008zz} hopes to reach $10^{-16}$. Similarly, the bound on $\tau \rightarrow \eta \mu$ should be reduced at future B-factories \cite{Akeroyd:2004mj}.

\section{Discussion}\label{conclu}
A natural way to approach the hierarchy and flavour problems is to assume that electroweak symmetry breaking arises via strong-coupling and that fermion masses arise via mixing of elementary fermions with composite fermions of the strong sector. The composite fermions carry both colour and electroweak charges and, if this is indeed how Nature works, it seems reasonable to at least entertain the possibility of the presence of composite bosons carrying both types of charge and acting as leptoquarks. Such states,  which couple predominantly to third-generation quarks and leptons, are even more plausible in the context of theories featuring unification of gauge couplings. The method of fermion mass generation gives rise to a natural suppression of leptoquark-mediated rare processes, and indeed my estimates, albethey very rough, suggest that the many constraints could be evaded. A more accurate calculation would require a concrete (and calculable) model. Models based on the AdS/CFT correspondence spring to mind, but given that such models already struggle to satisfy the more conventional flavour and electroweak precision constraints, the usefulness of searching for such a model is not obvious.

Typically, the leptoquarks sit at the TeV scale, but could be somewhat lighter if they arise as pseudo-Nambu-Goldstone bosons. Whether they do or not depends, of course, on the as-yet-unknown pattern of EWSB. In that case, the leptoquarks get a large, positive-definite contribution to their mass-squareds from QCD loop corrections, and end up with masses above the existing Tevatron bounds of {\em c.} 220 GeV.

Whilst leptoquark states are clearly not a necessity in this kind of scenario, their plausibility and consistency with existing constraints make them an obvious target for an LHC search, and the design of a suitable search strategy, ready for the advent of LHC data, would seem to be a priority. This should not be too difficult to achieve, given that the leptoquarks are within reach, that their production cross sections are roughly known, and that there are only four final states of interest, namely either $2t$ or $2b$, together with either $2\tau$ or $\slashed{E}_T$. If leptoquarks predominantly coupled to third generation quarks and leptons were to be discovered at the LHC, this would, in my opinion, provide a strong endorsement that the hierarchy and flavour problems are indeed solved in the framework of strong-coupling and partial fermion compositeness.
\begin{acknowledgments}
I am grateful to R.~Rattazzi for providing the motivation for this work, and to K.~Agashe and J.~Zupan for comments on the manuscript. I also thank A.~J.~Barr, B.~Campbell, C.~Grojean, A.~Pomarol, M~Redi, G.~G.~Ross, and A.~Weiler for discussions and the Les Houches Workshop on Physics Beyond the Standard Model for hospitality.
\end{acknowledgments}
\appendix
\section*{Appendix: Generators of $SU(4)$}
I choose a basis of generators $T^{i} = \frac{\lambda^{i}}{\sqrt{2}}$ in the fundamental of $SU(4)$, where $\lambda^{1\dots 8}$ are the Gell-Mann matrices supplemented with zeros in the last row and column and the rest are
\begin{widetext}
\begin{align}
\lambda^9 &= \begin{pmatrix} 0 & 0 & 0& 1\\ 0 & 0 & 0& 0 \\  0 & 0 & 0& 0 \\ 1 & 0 & 0& 0 \end{pmatrix}, 
\lambda^{10}= \begin{pmatrix} 0 & 0 & 0& -i\\ 0 & 0 &0& 0 \\  0 & 0 & 0& 0 \\ i & 0 & 0& 0 \end{pmatrix}, 
\lambda^{11} =\begin{pmatrix} 0 & 0 & 0& 0\\ 0 & 0 &0& 1 \\  0 & 0 & 0& 0 \\ 0& 1 & 0& 0 \end{pmatrix}, 
\lambda^{12}=\begin{pmatrix} 0 & 0 & 0& 0\\ 0 & 0 &0& -i \\  0 & 0 & 0& 0 \\ 0& i & 0& 0 \end{pmatrix},  \\
\lambda^{13}&=\begin{pmatrix} 0 & 0 & 0& 0\\ 0 & 0 &0& 0 \\  0 & 0 & 0& 1 \\ 0& 0 & 1& 0 \end{pmatrix}, 
\lambda^{14}= \begin{pmatrix} 0 & 0 & 0& 0\\ 0 & 0 &0& 0 \\  0 & 0 & 0& -i \\ 0& 0 & i & 0 \end{pmatrix}, 
\lambda^{15}= \frac{1}{\sqrt{6}} \begin{pmatrix} 1 & 0 & 0& 0\\ 0 & 1 &0& 0 \\  0 & 0 & 1& 0 \\ 0& 0 & 0 & -3 \end{pmatrix}.
\end{align}
\end{widetext}

\end{document}